# Identification of an Epidemiological Model to Simulate the COVID-19 Epidemic using Robust Multi-objective Optimization and Stochastic Fractal Search

**Gustavo Barbosa Libotte[1], Fran Sérgio Lobato[2] and Gustavo Mendes Platt[3]**

[1]Polytechnic Institute, Rio de Janeiro State University, Nova Friburgo, Brazil
[2]Chemical Engineering Faculty, Federal University of Uberlândia, Uberlândia, Brazil
[3]School of Chemistry and Food, Federal University of Rio Grande, Santo Antônio da Patrulha, Brazil

Correspondence should be addressed to Gustavo Barbosa Libotte; gustavolibotte@iprj.uerj.br

Emails: fslobato@ufu.br (Fran Sérgio Lobato), gmplatt@furg.br (Gustavo Mendes Platt)

**Abstract**

Traditionally, the identification of parameters in the formulation and solution of inverse problems considers that models, variables and mathematical parameters are free of uncertainties. This aspect simplifies the estimation process, but does not consider the influence of relatively small changes in the design variables in terms of the objective function. In this work, the SIDR (Susceptible, Infected, Dead and Recovered) model is used to simulate the dynamic behavior of the novel coronavirus disease (COVID-19), and its parameters are estimated by formulating a robust inverse problem, that is, considering the sensitivity of design variables. For this purpose, a robust multi-objective optimization problem is formulated, considering the minimization of uncertainties associated to the estimation process and the maximization of the robustness parameter. To solve this problem, the Multi-objective Stochastic Fractal Search algorithm is associated with the Effective Mean concept for the evaluation of robustness. The results obtained considering real data of the epidemic in China demonstrate that the evaluation of the sensitivity of the design variables can provide more reliable results.

## Introduction

Since the first case of coronavirus disease-2019 (COVID-19), a total of 82,877 confirmed cases and 4,633 deaths have been reported in China, up to May 2, 2020. This infectious disease may cause severe respiratory illness, fever, cough, and myalgia or fatigue are common symptoms at the onset of illness [7]. The transmission of this disease can be person-to-person, by direct contact with secretions or airborne droplets [4, 8, 15].

In order to study the dissemination of this disease, various mathematical models have been proposed. These models are based on compartments and relations between the different groups of individuals. Lin et al. [9] proposed a modification for the well-known Susceptible-Exposed-Infectious-Removed (SEIR) model for the COVID-19 outbreak in Wuhan, which considers some essential elements such as individual reaction, governmental actions, zoonotic transmission and emigration of a large portion of the population in a short period of time. Benvenuto et al. [2] proposed the Auto Regressive Integrated Moving Average (ARIMA) model to predict the spread, prevalence and incidence of COVID-2019. Roda et al. [16] used a Susceptible-Infectious-Removed (SIR) model to predict the COVID-19 epidemic in Wuhan after the lockdown and quarantine. In this study, the authors demonstrate that non-identifiability in model calibrations using the confirmed-case data is the main reason for such wide variations. Prem et al. [14] proposed a SEIR-like model to simulate the spread of COVID-19 in Wuhan city, where all demographic changes in the population (births, deaths and ageing) are ignored. This study shows that contingency plans aimed at reducing social contact can be effective in decreasing the magnitude and delaying the peak of the COVID-19 outbreak.

In all these models, no information about uncertainties are considered, that is, errors associated with modelling and estimation of parameters are neglected. In this case, optimal solutions may be influenced by small perturbations, as demonstrated by Deb and Gupta [6]. The objective of this work is to analyze the influence of noise during the solution of an inverse problem that aims to determine the parameters that characterizes a SIDR (Susceptible, Infected, Dead and Recovered) model. For this purpose, a robust inverse problem is formulated and solved using the Stochastic Fractal Search [17]. The problem proposed in this work considers the minimization of deviation between experimental and calculated values, together with the maximization of robustness parameter.

## Background Information on Robustness Analysis

In many optimization problems, decision variables are subject to perturbation. When solving the problem, one must consider that the solution must be acceptable with respect to small changes in the values of the decision variables. Robust optimization aims to obtain solutions that are least sensitive to such perturbations, that is, solutions that present the smallest possible deviation in relation to the objective value when subject to noise. The concepts of robust optimization were introduced by Tsutsui and Ghosh [19]. The objective function to optimize in searching robust solutions may be formulated as

$$f^R(x) = \int_{-\infty}^{+\infty} f(x+\delta)\, p(\delta)\, \mathrm{d}\delta,$$

where $\boldsymbol{\delta}$ is the noise parameter, and $\boldsymbol{p(\delta)}$ represents the probability distribution function. Usually, this effective objective function is not available, since the probability distribution may not be known. Therefore, the calculation of the expected performance is usually not trivial in many applications.

Deb and Gupta [6] proposed a methodology for obtaining robust solutions in the context of multi-objective optimization (more details on multi-objective optimization can be seen in Miettinen [10]). Essentially, it is proposed to optimize the mean effective objective values computed at a point by averaging the function values of a few samples in its vicinity, instead of optimizing the original objective functions. Thus, a solution $x^*$ is called a multiobjective robust solution of type I, if it is the global feasible Pareto-optimal solution to the multi-objective minimization problem given by

$$\begin{aligned}\text{Min}\quad & \left(f_1^{\text{eff}}(x), \ldots, f_m^{\text{eff}}(x)\right), \\ \text{Subject to}\quad & x \in \mathcal{S}\end{aligned}$$

where $\boldsymbol{f_i^{\text{eff}}(x)}$ is defined as

$$f_i^{\text{eff}}(x) = \frac{1}{|\mathcal{B}_\delta(x)|} \int_{y \in \mathcal{B}_\delta(x)} f_i(y)\, \mathrm{d}y, \qquad (1)$$

for $\boldsymbol{i = 1, \ldots, m}$, and $\boldsymbol{\mathcal{B}_\delta(x)}$ is the hypervolume of the neighborhood.

This approach is suitable for problems in which the result of the integral in Eq. (1) can be obtained in a closed analytical form. For problems in which the search space is more complex, Eq. (1) can be approximated by the mean value of the objective function, using a Monte Carlo integration given by

$$f_i^{\text{eff}}(x) \approx \frac{1}{H} \sum_{k=1}^{H} f_i(y^k).$$

In practice, a set of $\boldsymbol{H}$ points are randomly sampled (or respecting some structured manner, such as the Latin

Hypercube method) in the range $y^k \in [(1-\delta)x, (1+\delta)x]$, and the mean function value approximates Eq. (1).

## Stochastic Fractal Search

In the last decades, the development of optimization algorithms based on swarm intelligence has allowed the solution of complex problems in different areas of science and engineering. Inspired by the collective intelligent behavior of insects or animal groups in nature, such as flocks of birds, swarms of bees (bats, fireflies), colonies of ants, and swarms of fruit flies, various optimization algorithms have been proposed [18]. In this work, we use a promising method, recently proposed by Salimi [17], called Stochastic Fractal Search (SFS) algorithm. The main steps of the metaheuristic are presented below.

The SFS algorithm is a nature-inspired metaheuristic based on the natural growth phenomenon of fractals. Candidates solutions (particles) explore the search space considering a diffusion property which is regularly seen in random fractals. The SFS approach is based on random fractals grown by Diffusion Limited Aggregation concept [22]. This optimization strategy adopts a random walk in order to simulate the diffusion process, where the diffusing particle sticks to the seed particle. This process is repeated until a cluster has been created [17].

Two steps are applied to generate new candidate solutions at each iteration: diffusion and updating. In the first, each particle diffuses around its current position to ensure the exploitation property. The diffusion process avoids being trapped in local optimum and increases the chance of finding the global solution. In the second, a point in a group updates its location based on the locations of other points in the group. SFS considers a static diffusion process, that is, the best particle generated from the diffusion process is the only particle that is considered; the rest of the particles are ignored. In addition to efficient exploration of the feasible problem, SFS uses some random methods as updating processes.

The diffusion process uses Gaussian random walks to generate points around each particle until a predefined maximum diffusion number is reached. There are two types of Gaussian walks in the diffusion process:

$$G_1 = \text{Gaussian}(\mu_{BP}, \sigma) + (\varepsilon \times BP - \varepsilon' \times P_i)$$
$$G_2 = \text{Gaussian}(\mu_P, \sigma)$$

where $\varepsilon$ and $\varepsilon'$ are uniformly distributed random number in the range $(0, 1)$. In turn, $BP$ and $P_i$ are the position of the best point and the $i$-th point in the group, respectively. The first two Gaussian parameters are $\mu_{BP}$ and $\sigma$, where $\mu_{BP}$ is exactly equal to $BP$. The two the latter parameters are $\mu_P$ and $\sigma$, where $\mu_P$ is equal to $P_i$. The standard deviation $\sigma$ is dynamically adjusted based on the number of the generation $g$,

$$\sigma = \left|\frac{\log(g)}{g} \times (P_i - BP)\right|.$$

The update process employs two statistical procedures to undertake the exploration in SFS. Initially, all the points are ranked based on the value of the fitness function, by calculating

$$Pa_i = \frac{\text{rank}(P_i)}{N}, \qquad (2)$$

where $\text{rank}(P_i)$ is the rank of the point $P_i$ among the other points of the group. In the first updating process, for each point $P_i$ in group, the $j$-th component of $P_i$ is updated according to

$$P'_i(j) = P_r(j) - \varepsilon \times (P_t(j) - P_i(j)),$$

where $P_r$ and $P_t$ are randomly selected points in the group. The point is updated if the condition $Pa_i < \varepsilon$ is satisfied, where $Pa_i$ is given by Eq. (2). Otherwise, $P_i$ remains unchanged. In the second update process, $P'_i$ is updated if $Pa_i < \varepsilon$ holds for $P'_i$. Thus, the point is modified according to

$$P''_i = P'_i - \hat{\varepsilon} \times (P'_t - BP),$$

if $\varepsilon' \leq 0.5$, where $\varepsilon'$ is a random numbers generated by the Gaussian distribution. Otherwise, $P'_i$ is updated by

$$P''_i = P'_i + \hat{\varepsilon} \times (P'_t - P'_r).$$

In these cases, $P'_t$ and $P'_r$ are randomly selected points obtained from the first procedure. Details on the implementation are presented by Salimi [17].

## Multi-objective Stochastic Fractal Search

In this work, the SFS strategy is extended for multi-objective optimization context. This new approach, called the Multi-objective Optimization Stochastic Fractal Search (MOSFS) algorithm, incorporates two operators to the original SFS algorithm: Fast Non-Dominated Sorting and Crowding Distance [5]. Briefly, MOSFS presents the following structure. An initial population of size $NP$ is randomly generated. Then, a new population is generated from the current population, using the operators proposed in SFS. Each candidate of the new population is evaluated considering the vector of objectives. All dominated candidate solutions are removed from the population through the Fast Non-Dominated Sorting operator. The population is sorted into non-dominated fronts (sets of vectors that are non-dominated with respect to each other). This procedure is repeated until all vectors are assigned to a front. During the evolutionary process, if the number of individuals in the current population is larger than a predefined number, it is truncated according to the Crowding Distance.

## Methodology

Traditionally, models based on compartments have been used to represent dynamic behavior of disease. In the literature, various applications involving these type of model can be found. Among these, we can to cite studies involving: measles vaccination [1, 20], HIV/AIDS [11], tuberculosis [3], dengue [21], pertussis epidemiology [13], ebola virus [12], among others.

The objective of this work is to determine the parameters of an epidemiological model to predict the evolution of COVID-19 epidemic considering experimental data from China, considering uncertainties. For this purpose, the SIDR (Susceptible, Infectious, Dead and Recovered) model is adopted [12].

In this model, it is assumed that an individual/host is susceptible to infection, and the disease can be transmitted from any infected individual to any susceptible individual. The number of susceptible individuals varies with time according to

$$\frac{dS}{dt} = -\beta SI, \quad (3)$$

where $t$ is the time, and $\beta$ is the probability of disease transmission per contact. Any infected individual can transmit the disease to a susceptible or exposed individual. The number of infected individuals is calculated by

$$\frac{dI}{dt} = \beta SI - \frac{\gamma}{1-\rho} I, \quad (4)$$

where $\gamma$ is the per-capita recovery rate, and $\rho$ is the death rate. In turn, the number of dead individuals is calculated by

$$\frac{dD}{dt} = \frac{\rho}{1-\rho} \gamma I. \quad (5)$$

Once an individual has been moved from Infected to Recovered class, they cannot return to the Susceptible, Exposed, or Infected compartments class. Thus, the number of recovered individuals is obtained from

$$\frac{dR}{dt} = \gamma I. \quad (6)$$

The initial conditions of the system are given by $(S(0), I(0), D(0), R(0)) = (S_0, I_0, D_0, R_0)$.

In order to determine the SIDR parameters, it is required to formulate and solve an inverse problem. In general, the identification procedure consists in obtaining the model parameters through the minimization of the difference between calculated and experimental values. For the real data of the Epidemic in China, both infected ($I^e$) and dead ($D^e$) time series are known. Thus, the inverse problem can be defined as

$$F(I^c, D^c) = \frac{1}{\max(I^e)^2} \sum_{i=1}^{M} (I_i^e - I^c)^2 + \frac{1}{\max(D^e)^2} \sum_{j=1}^{N} (D_j^e - D^c)^2 . \qquad (7)$$

where $I^c$ and $D^c$ are the calculated values for infected and dead individuals, respectively. $M$ and $N$ are the number of experimental data for $I^c$ and $D^c$, respectively.

In order to evaluate the influence of perturbations in this (nominal) optimization problem, the robustness of the variables $I^c$ and $D^c$ is analyzed. In this case, for each candidate solution generated by using the MOSFS algorithm, $H$ points are sampled using Latin Hypercube method. Each point is evaluated considering the system of ordinary differential equations given by Eqs. (3)–(6). For this purpose, the Fourth-Order Runge-Kutta method is used. After such simulations, the Mean Effective approach is employed. Then, the mean effective objective is associated to the candidate generated by MOSFS. This procedure is performed until the maximum number of generations is reached.

## Results and Discussion

Considering the methodology presented, two cases are analyzed: in the first, the parameters of the SIDR model are estimated through the formulation and solution of an inverse problem without considering possible uncertainties; In the second, the influence of possible perturbations on the decision variables is considered. For this purpose, the criteria below are defined.

In both problems analyzed, the parameters of the SIDR model are defined in the following intervals (obtained after preliminary runs): $0.1 \leq \beta \leq 0.6$, $0.04 \leq \gamma \leq 0.6$, $0 \leq I_0 \leq 1$, and $0 \leq \rho \leq 1$. The results correspond to 20 runs of the metaheuristics (using a different seed for each run) for a maximum of 250 generations in each run. The initial conditions of the compartmental model are given by $(S(0), I(0), D(0), R(0)) = (1 - I_0, I_0, 0, 0)$, and the COVID-19 data are retrieved from Worldometer [23].

In the deterministic inverse problem, Eq. (7) must be minimized. For this purpose, SFS is employed with 25 individuals in the population. Table 1 presents the results for the best result and the standard deviation. It can be noted that the optimal value of $F$ is relatively high. However, the corresponding standard deviation value indicates that the metaheuristic obtained very close optimizers in all runs. This is due to the fact that the population dynamics undergoes many variations in the course of the epidemic. Such variations are due to government actions, disease mitigation strategy, and capacity of the health network. All of these aspects have an impact on the time series of the number of infected and dead individuals. Therefore, the best curve fitting is not necessarily able to describe the behavior of the epidemic at all times.

Table 1: Results of the deterministic inverse problem obtained in 20 executions of the SFS method, with $t_f = 95$ days.

| | $\beta$ (day$^{-1}$) | $\gamma$ (day$^{-1}$) | $I_0$ (−) | $\rho$ (day$^{-1}$) | $F$ (−) |
|---|---|---|---|---|---|
| Best | $3.686 \times 10^{-1}$ | $7.5125 \times 10^{-2}$ | $2.7807 \times 10^{-3}$ | $3.0866 \times 10^{-2}$ | 0.8249 |
| Standard deviation | $1.545 \times 10^{-8}$ | $1.7875 \times 10^{-8}$ | $1.2128 \times 10^{-7}$ | $2.3232 \times 10^{-8}$ | $1.9897 \times 10^{-9}$ |

Figure 1 presents the profiles considering the estimated parameters by solving the inverse problem defined by Eq. (7)—these results are weighted in relation to number of infected individuals, that is, the population size is a portion of the population that has been effectively tested. In Figs. 1(b) and 1(c) we can observe the accuracy of the curve fitting. Note that the time series present sudden changes between two consecutive points, that make the compartmental model not be able to describe its behavior more accurately. This is clear, mainly, from day 85 in the data referring to the number of dead individuals. Figure 3(a) presents the evolution of susceptible population during the epidemic. As expected, after the maximum value observed for the number of infected individuals (approximately between days 20 and 30), the number of susceptible individuals decreases.

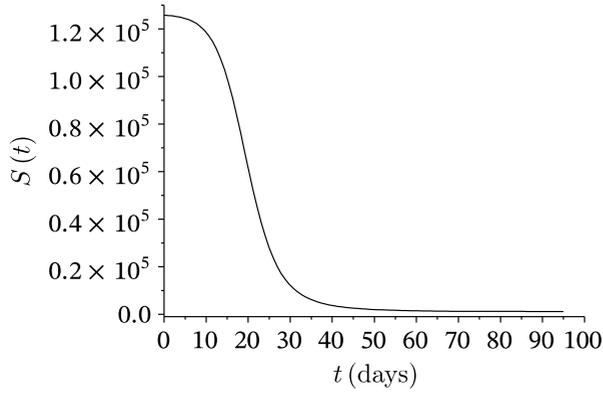
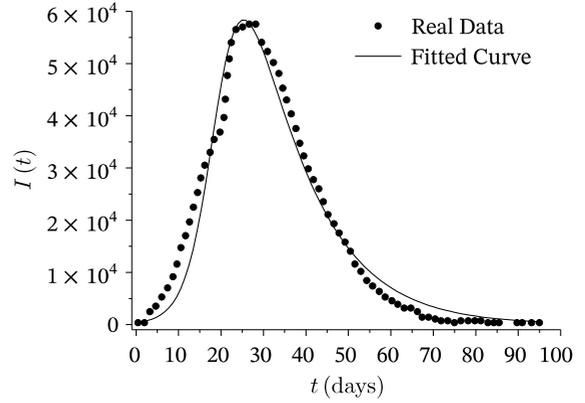

*(a) Susceptible individuals.*  *(b) Infected individuals.*

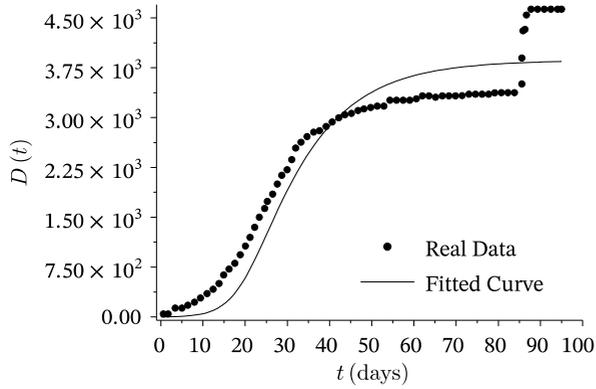
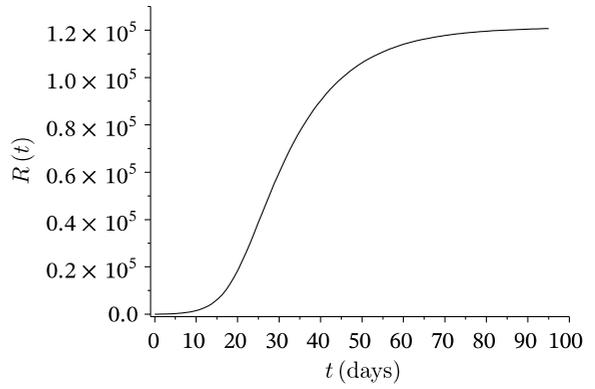

*(c) Dead individuals.*  *(d) Recovered individuals.*

*Figure 1: Profiles obtained by solving the inverse problem of Eq. (7). In the case of the classes of infected and dead individuals, the data used for the adjustment are also shown.*

In order to assess the influence of each parameter of the model on the objective function, the best solutions presented in Table 1 are perturbed, aiming to show the variation of $F$ in the vicinity of the optimal value of each parameter. The vector of optimal parameters is denoted by $\boldsymbol{\theta}^* = (\beta^*, \gamma^*, I_0^*, \rho^*)$. For each parameter, 100 equally spaced points are evaluated, in the range $[(1-\tau)\theta_k^*, (1+\tau)\theta_k^*]$, for $k = 1, \ldots, 4$, where $\tau = 0.25$. Figure 2 shows the influence of each parameter in relation to the objective function defined by Eq. (7). Note that, in the analyzed range, in fact, the optimal values of the corresponding parameters represent the minimum value of $F$. This indicates that, at least locally, the optimal parameters obtained represent the best fit in relation to the analyzed data. In addition, it is possible to notice that $\beta$ is the parameter that presents greater sensitivity in relation to $F$, in the proposed analysis.

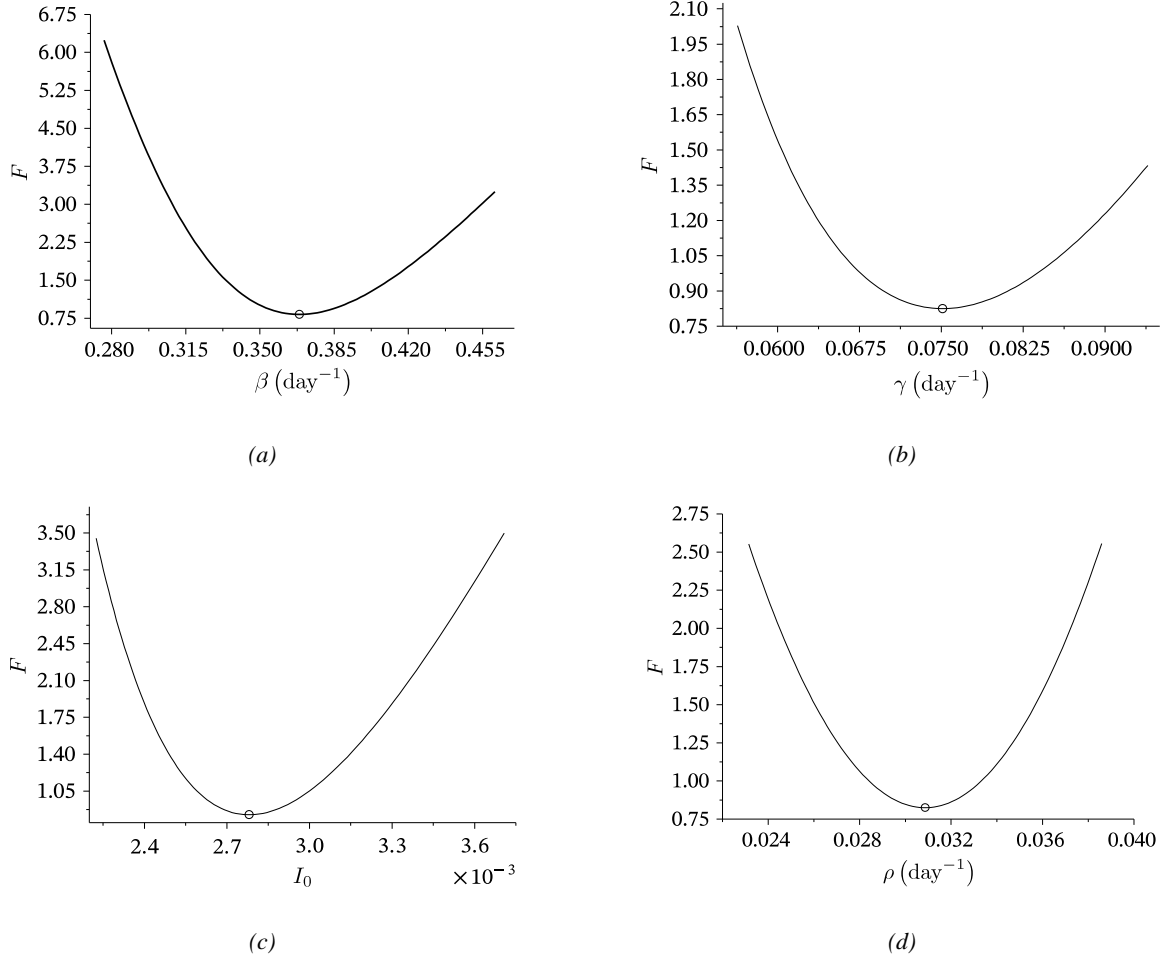

*Figure 2: Sensitivity analysis of (a) β; (b) γ; (c) $I_0$; (d) ρ in relation to the value of the objective function.*

In order to analyze the sensitivity of the design variables, a multi-objective optimization problem is formulated. In the robust inverse problem, Eq. (7) must be minimized and the noise parameter (δ) is maximized. By solving this new problem, it is possible to assess the parameters of the compartmental model in the presence of uncertainty. Thus, the adjustment of the time series of data considers possible variations caused by external factors, such as underreporting of infected and dead individuals. Figure 3 shows the Pareto curve obtained for the robust inverse problem, calculated by the MOSFS method, using the Effective Mean approach with $H = 50$ random samples, noise parameter varying in the range $0 \leq \delta \leq 0.1$, and the same parameters adopted in the previous problem.

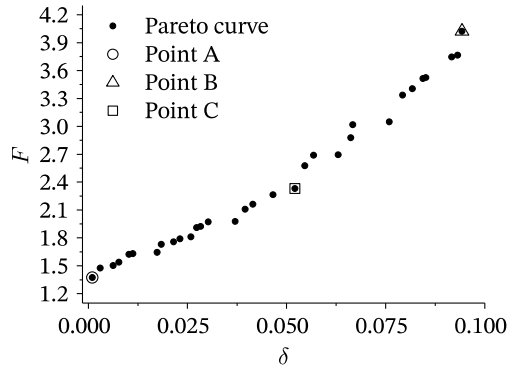

*Figure 3: Pareto curve of the robust inverse problem.*

The analysis of Fig. 3 shows that, as the noise parameter δ increases, the value of the objective $F$ also increases. Initially, note that the nominal solution is equivalent to the point on the Pareto curve that $\delta = 0$ (see Table 1). Increasing $F$ means that the associated fitted curve departs from the actual data as δ increases. Such displacement is caused by noise in the design variables, which can be understood as the existing uncertainties in the real data. Table 2 presents some highlighted points of the Pareto curve (points A, B, and C), as shown in Fig. 3.

*Table 2: Some points of the robust inverse problem obtained with MOSFS method. Points A, B, and C are highlighted in Fig. 3.*

| Point | $\delta$ | $\beta$ (day$^{-1}$) | $\gamma$ (day$^{-1}$) | $I_0$ (−) | $\rho$ (day$^{-1}$) | $F$ (−) |
|---|---|---|---|---|---|---|
| A | $9.8432 \times 10^{-4}$ | $3.7746 \times 10^{-1}$ | $5.9720 \times 10^{-2}$ | $2.8831 \times 10^{-3}$ | $3.2556 \times 10^{-2}$ | 1.3739 |
| B | $9.4248 \times 10^{-2}$ | $3.9105 \times 10^{-1}$ | $4.5508 \times 10^{-2}$ | $2.9617 \times 10^{-3}$ | $3.6130 \times 10^{-2}$ | 4.0218 |
| C | $5.2114 \times 10^{-2}$ | $3.8435 \times 10^{-1}$ | $4.9799 \times 10^{-2}$ | $3.0543 \times 10^{-3}$ | $3.3932 \times 10^{-2}$ | 2.3307 |

Figure 4 shows the profiles corresponding to the parameters of points A, B, and C, shown in Table 2. Each of these three points presents a different compromise in relation to the objectives of the robust inverse problem. Point A, represented by a circle, presents an extreme compromise in relation to F, that is, it does not prioritize robustness. On the other hand, point C, represented by a diamond, has an extreme compromise to robustness. In turn, point B, represented by a triangle, presents an intermediate compromise between both objectives. Figure 4 also shows the profile which is equivalent to the nominal inverse problem solution ($\delta = 0$), represented by the solid line.

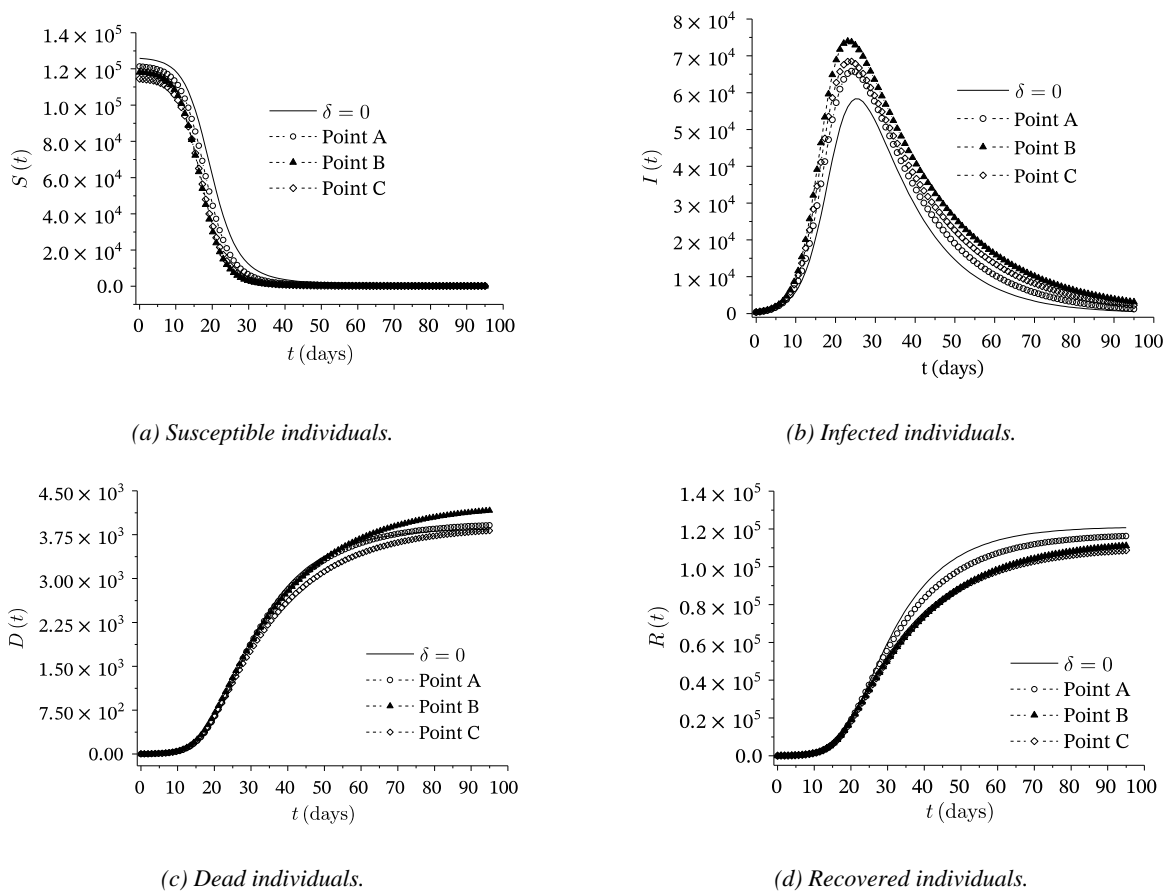

*(a) Susceptible individuals.*  *(b) Infected individuals.*

*(c) Dead individuals.*  *(d) Recovered individuals.*

*Figure 4: Profiles obtained by solving the robust inverse problem with different values for δ. Points A, B, and C denote the profiles corresponding to the parameters listed in Table 2. The profile represented by the solid line corresponds to the nominal case, whose parameters are shown in Table 1.*

Especially in relation to the curve of infected people as a function of time (Fig. 4b), note that the solution of the robust inverse problem provides profiles which are shifted in relation to the result corresponding to the nominal case. Such displacements tend to become more pronounced when the required level of robustness increases. Thus, the profiles of the susceptible, dead, and recovered compartments are also shift, in order to maintain $N(t) = S(t) + I(t) + D(t) + R(t)$ constant.

## Conclusions

In this work we proposed and solved two inverse problems (nominal and robust) to simulate the dynamic behavior of COVID-19 considering real data from China. Stochastic Fractal Search algorithm was employed to solve the nominal inverse problem. We also proposed an extension of SFS in the multi-objective context to solve the robust inverse problem. In order to assess uncertainties, the mean effective concept was considered. The parameters of the compartmental SIDR model were determined and analysed considering both nominal and robust context. In order to analyse the influence of uncertainties, a multi-objective optimization problem was formulated and solved. This problem considers the minimization of deviations associated to the experimental data and calculated values considering the proposed model and the maximization of the robustness parameter. In general, the solution of the proposed multi-objective problem demonstrates that the increase in the noise parameter implies an increase in the value of the objective function. The use of the proposed robust approach to estimate the compartmental model parameters demonstrates the importance of incorporating a methodology to assess the robustness during the solution of the proposed inverse problem. Finally, the use of a mathematical model associated with optimization tools may contribute in the future to the study and development of strategies to understand the dynamic behavior of COVID-19.

## References

[1]     C. T. Bauch, E. Szusz, and L. P. Garrison, "Scheduling of Measles Vaccination in Low-income Countries: Projections of a Dynamic Model," *Vaccine*, vol. 27, no. 31, pp. 4090–4098, Jun. 2009, doi: 10.1016/j.vaccine.2009.04.079.

[2]     D. Benvenuto, M. Giovanetti, L. Vassallo, S. Angeletti, and M. Ciccozzi, "Application of the ARIMA Model on the COVID-2019 Epidemic Dataset," *Data Br.*, vol. 29, p. 105340, Apr. 2020, doi: 10.1016/j.dib.2020.105340.

[3]     S. Bowong and J. Kurths, "Parameter Estimation Based Synchronization for an Epidemic Model with Application to Tuberculosis in Cameroon," *Phys. Lett. A*, vol. 374, no. 44, pp. 4496–4505, Oct. 2010, doi: 10.1016/j.physleta.2010.09.008.

[4]     J. F.-W. Chan *et al.*, "A Familial Cluster of Pneumonia Associated with the 2019 Novel Coronavirus Indicating Person-to-person Transmission: a Study of a Family Cluster," *Lancet*, vol. 395, no. 10223, pp. 514–523, Feb. 2020, doi: 10.1016/S0140-6736(20)30154-9.

[5]     K. Deb, A. Pratap, S. Agarwal, and T. Meyarivan, "A Fast and Elitist Multiobjective Genetic Algorithm: NSGA-II," *IEEE Trans. Evol. Comput.*, vol. 6, no. 2, pp. 182–197, Apr. 2002, doi: 10.1109/4235.996017.

[6]     K. Deb and H. Gupta, "Introducing Robustness in Multi-objective Optimization," *Evol. Comput.*, 2006, doi: 10.1162/evco.2006.14.4.463.

[7]     C. Huang *et al.*, "Clinical Features of Patients Infected with 2019 Novel Coronavirus in Wuhan, China," *Lancet*, vol. 395, no. 10223, pp. 497–506, Feb. 2020, doi: 10.1016/S0140-6736(20)30183-5.

[8]     Q. Li *et al.*, "Early Transmission Dynamics in Wuhan, China, of Novel Coronavirus–Infected Pneumonia," *N. Engl. J. Med.*, vol. 382, no. 13, pp. 1199–1207, Mar. 2020, doi: 10.1056/NEJMoa2001316.

[9]     Q. Lin *et al.*, "A Conceptual Model for the Coronavirus Disease 2019 (COVID-19) Outbreak in Wuhan, China with Individual Reaction and Governmental Action," *Int. J. Infect. Dis.*, vol. 93, pp. 211–216, Apr. 2020, doi: 10.1016/j.ijid.2020.02.058.

[10]  K. Miettinen, *Nonlinear Multiobjective Optimization*, 1st ed. New York, New York, USA, 1998.

[11]  Z. Mukandavire, C. Chiyaka, W. Garira, and G. Musuka, "Mathematical Analysis of a Sex-structured HIV/AIDS Model with a Discrete Time Delay," *Nonlinear Anal. Theory, Methods Appl.*, vol. 71, no. 3–4, pp. 1082–1093, Aug. 2009, doi: 10.1016/j.na.2008.11.026.

[12]  A. C. Osemwinyen and A. Diakhaby, "Mathematical Modelling of the Transmission Dynamics of Ebola Virus," *Appl. Comput. Math.*, vol. 4, no. 4, p. 313, 2015, doi: 10.11648/j.acm.20150404.19.

[13]  P. Pesco, P. Bergero, G. Fabricius, and D. Hozbor, "Modelling the Effect of Changes in Vaccine Effectiveness and Transmission Contact Rates on Pertussis Epidemiology," *Epidemics*, vol. 7, pp. 13–21, Jun. 2014, doi: 10.1016/j.epidem.2014.04.001.

[14]  K. Prem *et al.*, "The Effect of Control Strategies to Reduce Social Mixing on Outcomes of the COVID-19 Epidemic in Wuhan, China: a Modelling Study," *Lancet Public Heal.*, Mar. 2020, doi: 10.1016/S2468-2667(20)30073-6.

[15]  J. Riou and C. L. Althaus, "Pattern of Early Human-to-human Transmission of Wuhan 2019 Novel Coronavirus (2019-nCoV), December 2019 to January 2020," *Eurosurveillance*, vol. 25, no. 4, Jan. 2020, doi: 10.2807/1560-7917.ES.2020.25.4.2000058.

[16]  W. C. Roda, M. B. Varughese, D. Han, and M. Y. Li, "Why is it Difficult to Accurately Predict the COVID-19 Epidemic?," *Infect. Dis. Model.*, vol. 5, pp. 271–281, 2020, doi: 10.1016/j.idm.2020.03.001.

[17]  H. Salimi, "Stochastic Fractal Search: A Powerful Metaheuristic Algorithm," *Knowledge-Based Syst.*, vol. 75, pp. 1–18, Feb. 2015, doi: 10.1016/j.knosys.2014.07.025.

[18]  E.-G. Talbi, *Metaheuristics*. Hoboken, NJ, USA: John Wiley & Sons, Inc., 2009.

[19]  S. Tsutsui and A. Ghosh, "Genetic Algorithms with a Robust Solution Searching Scheme," *IEEE Trans. Evol. Comput.*, vol. 1, no. 3, pp. 201–208, 1997, doi: 10.1109/4235.661550.

[20]  P. Widyaningsih, D. R. S. Saputro, and A. W. Nugroho, "Susceptible Exposed Infected Recovery (SEIR) Model with Immigration: Equilibria Points and Its Application," in *AIP Conference Proceedings*, 2018, doi: 10.1063/1.5054569.

[21]  H. H. Weiss, "The SIR Model and the Foundations of Public Health," *Mater. MATemàtics*, vol. 2013, no. 3, pp. 17, 2013.

[22]  T. A. Witten and L. M. Sander, "Diffusion-limited Aggregation," *Phys. Rev. B*, vol. 27, no. 9, pp. 5686–5697, May 1983, doi: 10.1103/PhysRevB.27.5686.

[23]  Worldometer, "Worldometer's COVID-19 Data," 2020. https://www.worldometers.info/coronavirus/country/china/ (accessed May 16, 2020).